\documentclass[a4paper,12pt]{article}
\usepackage[colorlinks]{hyperref}
\usepackage{graphicx}
\begin{document}
\hypersetup{
  colorlinks   = true, 
  urlcolor     = blue, 
  linkcolor    = blue, 
  citecolor   = blue 
}
\thispagestyle{empty}
\begin{center}
    {\large\bf On some EPR (Einstein, Podolsky, Rosen) issues\\}
    \vskip3mm
   {\bf Giuseppe Giuliani\\}
   \vskip3mm
 { \small Formerly at: Dipartimento di Fisica, Universit\`a degli Studi di Pavia. Retired.}
   \vskip3mm
 {\small email: giuseppe.giuliani@unipv.it }
 \vskip3mm
 {\small website: \url{www.fisica.unipv.it/percorsi/}}
    \end{center}

\noindent{\bf Abstract.} A critical reconsideration of the EPR (Einstein - Podolsky - Rosen) paper shows  that the  EPR argument can be developed without using the concept of `element of physical reality', thus eliminating any
      philosophical element in the logical chains of the paper.
      Deprived of its philosophical ornament, the EPR argument plainly
      reduces to require what quantum mechanics can not do: to assign
      definite values to two incompatible physical quantities.
      Hidden variables theories built up according to Bell - type theorems are formulated on the basis of  the assumption that the locality condition implies the statistical independence between two measurements space - like separated.  This assumption is valid only with the additional one that statistical dependence between two measurements requires a causal connection between them. This additional assumption rules out the possibility that  statistical dependence may due to an intrinsic property of the physical system under study. Therefore, hidden variables theories are built up with a  restriction which leads them to be disproved by experiment. They appear to be `straw man' theories whose main role is that of putting under fire philosophical realism. However, a philosophy can not be disproved by an experiment unless it is shown that this experiment disproves a  theory whose postulates are  logical consequence of the philosophy.
   Quantum mechanical non - locality, invoked for describing EPR - type experiments, is strictly connected to the hypothesis (NDV hypothesis) according to which the twin photons of entangled pairs do not have a definite polarization before measurements. Both hypotheses are used only for describing EPR experiments and not for making predictions. Therefore, they can be dropped without reducing the predictive power of quantum mechanics concerning entangled photons pairs.  Furthermore,  both  hypotheses
      can be experimentally tested by a  modification of a standard experimental
      apparatus designed for studying entangled photons pairs.
\tableofcontents
\section{Introduction}
In 1935, Albert Einstein, Boris Podolsky and Nathan Rosen (EPR) published a paper in which the completeness of quantum mechanics was called into question \cite{eprP}. In a historical perspective, the origin of the paper must be viewed in the contest of the epistemological and philosophical debate that accompanied the foundation of quantum mechanics.
\par
Broadly speaking, the philosophical inspiration of the EPR paper is a realist one and, apparently, the EPR argument hinges upon the concept of `physical reality'  (section \ref{eprpaper}). This choice has mixed up epistemological considerations about the completeness of a theory with more general philosophical stands, thus entangling in an intricate way, physics, epistemology and philosophy. The literature about EPR issues is huge: for instance, the review paper on `Bell non - locality' \cite{cripto}, essentially dedicated to the physical offsprings of the EPR paper, counts over five hundred references; while  the site \href{http:www.philpapers.org}{philpapers.org}, returns 170 philosophical papers at the query  `Einstein Podolsky Rosen' and other 664 papers on related issues. In this situation, it is almost a foolhardy idea to enter the debate. Nonetheless, I dare to submit some considerations on three specific EPR issues: the EPR paper, the role of locality in Bell - type theorems and
 the concept of non - locality.
\par
  In   section \ref{philback} the philosophical background of the present paper will be expounded. It will be used for analyzing the philosophical content of the EPR paper and for bringing to light its role in the EPR argument. This  background will further be used for disentangling philosophy from physics in much more recent papers. The discussion of the EPR paper (section \ref{eprpaper}) will  show that the concept of `element of physical reality', apparently so  substantial to the paper, can be dropped without weakening the EPR argument. Section \ref{bohmsec} is dedicated to the Bohm's version of the thought experiment devised by EPR and to a summary of  the ensuing developments. The role of locality  condition in Bell's - type theorems is discussed in section \ref{belltype}. In the following section \ref{orsaysec}, a typical experiment with entangled photon pairs (Orsay, experiment, 1982) is presented and its standard interpretation discussed. In section \ref{howto},  it is shown how to experimentally test the two  hypotheses used in this standard interpretation: the hypothesis that the photons of an entangled pair do not have a definite value of  polarization before measurement (NDV hypothesis, Not Definite Value) and the subsequent necessary hypothesis of non - locality. Finally, the last section \ref{then} is dedicated to some recent entanglements of physics and philosophy in EPR issues.
\section{Philosophical and epistemological back\-ground of this paper\label{philback}}
In this section -- for sake of clarity --  I  outline the philosophical and epistemological positions that inspire this paper. These positions play a heuristics role, in the sense that they suggest how to deal with some philosophical EPR issues. But -- of course -- the theses held in the present paper, can not be logically derived from these philosophical and epistemological positions \footnote{A more detailed exposition of the epistemological stands of the present paper can be found in \cite[pp. 1 - 25]{erq}. Furthermore, in this reference, these epistemological positions are tested by applying them to fundamental turning points in the history of physics.}.
\par
The historical development of Science,  broadly  suggests that experimental disciplines have developed on the basis of three main philosophical
  assumptions:
 \begin{enumerate}
    \item There is  a World whom the observer belongs to.\label{world}
    \item Causality principle.\label{causality}
    \item  The World behaves constantly in the same way (phenomena are reproducible).\label{constancy}
 \end{enumerate}
 The epistemic status of these assumptions is different. The first is a
 reasonable hypothesis and, though it has been challenged with various argumentations, often paradoxical, it constitutes the cornerstone of experimental inquiry.
  The second has proved to be  of great heuristic value because it asks for a methodological search of the causes of phenomena.
 The third is a necessary  prerequisite for any knowledge of the World and has been sustained by thousands of years  of observations and experimental inquires.  These
 same assumptions constitute the foundations
  of a rationally oriented common sense
 and guide us in our daily life.
  \par
 Assumption (P1) asks for an answer to a basic question: which are the
 relationships between scientific descriptions of the World (or part of it) and
 the  World? Answering this question amounts to  sketch a `theory of scientific
 knowledge': whatever it would be, {\em this theory should be
 independent from the various disciplines, their  theories
 and experimental results}.
 \par
We shall try to answer in some degree this question with reference to Physics: to
see if and how the following considerations apply also to other
experimental disciplines goes beyond the scope of this paper and of my capacities.
\subsection{Theoretical entities and physical quantities}
The descriptions of Physics in their mature stage come in as
theories which use~-~among others~-~two  basic types of concepts:
theoretical entities and physical quantities. Examples of theoretical entities
are the concepts  of reference frame, particle, wave,  electron, proton\dots
Physical quantities describe properties of theoretical entities or relations between them. For
instance, mass, charge, spin and magnetic moment describe properties
of the theoretical  entity `electron'. The physical quantity `velocity' is attributed to, for instance,  an electron with respect to a reference frame. The physical quantity `force' describes an interaction between two theoretical entities, for instance, between two masses. Physical quantities enter equations of Physics and can be
measured.
\subsection{Measurements and ontological statements}
 A measurement
 can be defined as a set of experimental procedures that allow to
 attribute to a physical quantity a definite value (within a range of
 experimental inaccuracy). The result of a
 measurement of the quantity $G$ that describes a property
 of the   theoretical entity $E$ depends on the interaction between the
 theoretical entity $E$ and the apparatus $A$ (also considered as a
   theoretical entity):  the result of the
 measurement  {\em depends} on the property of the  theoretical entity $E$ described by the
 quantity $G$.
 \par
As an example, let us consider  the  measurement of the mass of an
ion with a mass spectrometer: the outcome of the measurement depends
on a property (which we call `mass') of the ions we are testing. In
this  case, our acquired knowledge suggests that the apparatus does
not influence the result of the measurement. However, this is not,
in general, the case. For instance, the insertion of an ammeter in
an electrical circuit changes its electrical resistance and,
therefore, the measured value of the current is different from that
of the circuit without the ammeter.
\par
On the basis of assumption (P1), we state that the result of the
measurement reflects a property $P_{Q_E}$ of a {\em quid} $Q_E$
that, in the World, corresponds to the   theoretical entity $E$.
   We can only establish a
correspondence between   theoretical entities and {\em quid} and
properties of theoretical entities and properties of {\em quid}. For
instance, we can {\em say} that in the World there {\em is} a {\em
quid} that corresponds to our theoretical entity `electron': this
means that in the World there is a {\em quid} which has properties
that correspond to the properties attributed by our theory to the
`electron' and that this {\em quid} behaves in accordance with the
laws of our theory and with properties that are {\em described} by
the measured values of the quantities $G_i$ that our theory
attributes to the `electron'. We can  convene that the statement
`the electron exists in the World' is {\em simply and only} a {\em
shorthand} of the previous one.
\par
Ontological statements, like the previous one about the existence of
the electron, can be made only {\em a posteriori} by looking at our
entire acquired knowledge. They can not be deduced by logical chains
from the acquired knowledge, but they must be compatible with it:
 ontological statements can be only plausible.
    While nowadays the statement `the electron
exists' (as a shorthand of the longer one given above) is plausible,
the statement `the Ether exists' cannot be reasonably considered as
compatible with our present acquired knowledge. However, the existence of the Ether was plausible in Maxwell's times.
If the plausibility of an ontological statement holds up for a long time, it turns into likelihood. For instance, around 1900, the electron  was endowed with a mass and an electric charge, but its role in the constitution of matter was uncertain if not obscure: its existence was loosely plausible. Later, it acquired  an intrinsic angular momentum (spin) and an intrinsic magnetic moment. Also a wavelength has been associated to it: the de Broglie wavelength $\lambda_{dB}=h/p$. However, the physicists, who increasingly learned how to manipulate the `electrons' in spite of their changing properties, wisely kept on  referring to them as to the same `particle'. Meanwhile, the theory of the electron has changed  and we can not exclude that it will change again. But the statement of the existence of the electron has become more and more plausible: it has become probable.
 The statement of existence of a theoretical entity may become more stable than the theory which originally used it.
 \par
We can measure physical quantities  which describe properties of theoretical entities that do not exist. For instance, Maxwell, in the item {\em Ether}, included in the IX edition of the Enciclopedia Britannica, showed how to measure the rigidity of the Ether starting from the measurement of the intensity of light coming from the Sun \cite{etere}. This is possible because the theories themselves indicate or suggest what to measure. Another, more intriguing, example is given by  the `hole', a concept developed in  solid state physics in the late Twenties of last century. Theoretically, a `hole' is constituted by a vacant site in the valence band of a semiconductor: the behavior of the remaining electrons under the action of electric or magnetic fields can be described as due to the motion of a `particle', the hole, endowed with electric charge $+e$, (effective) mass and electric mobility. Solid state physicists measure these quantities, which are properties of the theoretical entity `hole'. But we can hardly say that the `hole' exists,  as we can say that the `electron'  exists  \cite{what}.
\par
We shall define   `realism' a philosophical position  who accepts the three postulates above. It is
 a loose definition that allows many versions.
We shall define   `tempered realism' a realism which incorporates the
analysis given above of the process of measurement with its
implications; in particular, a `tempered realism' agrees with the
above definition and use of ontological statements. A `tempered
realist' rejects naive
  assumptions. Let us consider a theory that, like classical
    electromagnetism,
       is in good agreement with experiments (in its domain of
    application).
   We conclude that  the phenomena
    described by classical electromagnetism happen in the World {\em exactly} as
     described by the theory. This conclusion implies that all the theoretical entities used by the theory exist in the World with the properties describes by the associated physical quantities. But we have seen that ontological statements are only plausible or verisimilar; furthermore, we should know, independently from our theory,
     that in the World things happen {\em exactly} as described by our theory.
     This kind of naive realism, that can be denoted as
    `realism of theories', appears as untenable.
    \par
    In dealing with the EPR paper and some EPR issues, we shall encounter many versions of realism. We shall also find out that, somewhere, it is  held that experiments can falsify a philosophy. This can be true only in one case: when the postulates of a theory are logical consequences  of a philosophy and the theory is disproved by an experiment. To my knowledge, in the history of physics there is not an example  of this kind and, reasonably, it never will there be.
   \section{The EPR paper\label{eprpaper}}
 The premise of the paper is a philosophical one:
 \begin{quote}
 Any serious consideration of a physical
theory must take into account the distinction between the {\em objective reality, which is independent of any theory}, and the physical concepts with which the theory operates. These concepts are intended to correspond with the objective reality, and by means of these concepts we picture this reality to ourselves (my italics) \cite[p. 777]{eprP}.
\end{quote}
  Between the objective reality and its physical descriptions there is a correspondence, but EPR do not clarify the properties of this correspondence. In the words of the philosophical background of section \ref{philback}, EPR hold that there is a World (the objective reality), i.e. they accept the supposition \ref{world} of page \pageref{world}.
 \par
Without defining what is meant by `physical reality', EPR formulated a sufficient condition ($SC$):
\begin{quote}\label{SC}
If, without in any way disturbing a system, we can predict with certainty (i.e., with probability equal to unity) the value of a physical quantity, then there exists an element of physical reality corresponding to this physical quantity \cite[p. 777]{eprP}.
\end{quote}
At a first sight, it is not clear if the `physical reality' coincides with the `objective reality'  or if the `physical reality' belongs to a `physical picture of the World' that must be distinguished from the `objective reality'.
   In the first case, the sufficient condition ($SC$) states that  an algorithm will allow to establish the existence in the World of a physical quantity, and, therefore, also of the theoretical entity of which the physical quantity describes a property.
   This interpretation is incompatible with the concept of objective reality (the World with its properties as a datum, distinguished from the physical descriptions of it), because some basic properties of the World will be established directly by an algorithm. Therefore, we are left with the latter case: {\em the concept of physical reality belongs to a physical description of the World}.
 \par
The sufficient condition -- as it stands -- leads to paradoxical implications. For example.
  If a photon has passed through a linear polarizer whose optical axis is aligned along the $z$ axis, the photon -- after the polarizer -- is linearly polarized along the $z$ axis. Then, we can predict with certainty, i.e. with probability equal to one, that the photon will pass through a second polarizer with it axis aligned along the $z$ axis. Therefore -- according to the sufficient condition ($SC$) -- the linear polarization of the photon is an element of  physical reality. Instead, if the axis of the second polarizer makes  an angle $\theta$ with the $z$ axis, we can predict only that the photon will pass through the second polarizer with probability $\cos ^2 \theta$ (Malus' law). In this case, the sufficient condition is not satisfied: therefore, the linear polarization is not an element of  physical reality. Then, being an element of  physical reality depends on the type of measurement we are planning to do: this result is hardly acceptable.
 Rephrasing this conclusion with the words of EPR: ``No reasonable
definition of reality could be expected to permit
this \cite[p. 780]{eprP}.
In order to overcome this embarrassing conclusion, the sufficient condition should be reformulated with the specification  that  it must be satisfied  in {\em at least} one physical situation. In this way, the linear polarization of the photon will be an element of the physical reality in both cases.
\par
 Again, without defining what is meant by `complete' when speaking of a theory,
EPR formulated
 a necessary condition ($NC$):
 \begin{quote}
 Every
element of the physical reality must have a counterpart in the physical theory \cite[p. 777]{eprP}.
 \end{quote}
 This formulation is ambiguous. It could be interpreted by saying that every element of physical reality must be quantitatively described in a complete theory. But this interpretation does not allows to demonstrate the EPR thesis. For doing so, it must be interpreted by stating that a complete theory must attribute -- in every physical situation -- a definite value to a physical quantity endowed with physical reality by the sufficient condition.
 \par
 There is a tension between the sufficient  and the necessary condition. While the sufficient condition must be satisfied  at least in one  physical situation, the necessary condition must be satisfied in every physical situation. Then, it is easier to find out physical quantities endowed with physical reality than to build up a complete theory that describes these physical quantities. Finally, though devised for quantum mechanics, these conditions should be applicable also to classical physics; otherwise, it should be explained why these two descriptions of the objective reality (or part of it)  must obey different epistemological criteria.
 \par
After these definitions,  EPR considered a system composed of two parts. The two parts are permitted to
interact from
 $t=0$ to $t=T$; for $t>T$ no further interaction between the two parts is
possible. If A and B
  are two physical quantities, when the two parts no longer interact, a measurement
of the quantity
   A on part 1 will leave part 2 in a definite state, say $\psi_2$ (reduction
of the wave
   function). If, instead,  the quantity B is measured on part 1, part 2 will
be left in a
   different state, say $\varphi_2$.  It may happen that $\psi_2$ and
$\varphi_2$ are
   eigenfunctions of two non - commuting operators corresponding to some
physical quantities
   P and Q. Therefore, without in any way perturbing 2, we can predict with
certainty the values
    of the non - commuting quantities P and Q of part 2.
Then, the quantities P and Q of part 2 are `elements of physical reality': as such, they should
 be described by a complete theory. But quantum mechanics says that two incompatible quantities P and Q can
not be described by a common eigenfunction; therefore, quantum mechanics can not attribute definite values to both P and Q. In this precise, technical sense, P and Q can not have
 simultaneous definite values. Therefore, quantum mechanics is incomplete.
 \par
 As an example, EPR  applied this argument to the case
of two
  particles described by a particularly chosen  wavefunction in one dimension:
the difference
   $x_1-x_2$ and the sum $p_{1_x}+p_{2_x}$ have definite values, $x_0$ and $0$,
respectively.
    This is permitted by the formalism because the quantities $x_1-x_2$ and
$p_{1_x}+p_{2_x}$
     are compatible, i.e. their operators commute. For this system, if a
measurement of $x_1$
      yields $x_1=x_m$, then we know  with certainty that a measurement of $x_2$
will yield
       $x_2=x_m+x_0$. On the other hand, if a measurement of $p_{1_x}$ yields
$p$, we know
        with certainty that a measurement of $p_{2_x}$ will yield $-p$. Hence,
both $x_2$
         and $p_{2_x}$ are elements of physical reality. But quantum mechanics
says that
         $x_2$ and $p_{2_x}$ can not have simultaneous definite values.
Consequently, quantum
         mechanics is incomplete because it does not account for all elements
of physical
         reality.
 \par
 However,
   EPR stress that:
   \begin{quote}
   Indeed, one would not arrive
at our conclusion if one insisted that two or more physical quantities can be regarded as simultaneous elements of reality {\em only when they can be simultaneously measured or predicted} (original italics). On this point of view, since either one or the other, but not both simultaneously, of the quantities $P$ and $Q$ can be predicted, they are not simultaneously real. This makes the reality of $P$ and $Q$ depend upon the process of measurement carried out on the first system, which does not disturb the second system in any way. No reasonable definition of reality could be expected to permit this  \cite[p. 780]{eprP}.
\end{quote}
As we have seen, the EPR argument is developed starting from  the definition of the `element of physical reality'. However, this concept is superfluous and obscures the intimate nature of the EPR argument.
In fact, it is possible to restate {\em both} EPR conditions in the following way:
         \begin{quote}
        If  we can predict with certainty
(i.e., with probability equal to one) the value of a physical quantity in at least one physical situation, then a complete theory must allow this quantity to have, {\em in every physical circumstance}, a definite value.
\end{quote}
 Clearly, this conditions can not be satisfied by quantum mechanics, when a couple
of incompatible
 quantities
like position and momentum are considered. Then, quantum mechanics
 is incomplete. This necessary reformulation of the EPR argument  reduces to plainly requiring what quantum mechanics can not do: to assign  definite values to two incompatible quantities.
  \par
The above analysis implies that the value of the EPR paper must be reconsidered. It is not a definite proof that quantum mechanics is incomplete, because its incompleteness is founded on a questionable definition of completeness. In fact,  it can be  held that other criteria  are important in the evaluation of a theory: is the theory internally coherent? its predictions are corroborated by experiment? it describes quantitatively all phenomena that reasonably fall in its application domain? Historically, these criteria have decided the fate of physical theories. Quantum mechanics can not be an exception. Of course, a theory can be challenged also on the basis of philosophical considerations. However, in this case, the
thing to do, is to develop an equivalent or better theory, i.e. a theory with an application domain equal to or larger than that of quantum mechanics.
\par
 In the EPR paper, the concept of locality never appears explicitly. EPR speak only of two parts of a system that no longer interact:
 \begin{quote}\small\label{eprloc}
    On the other
hand, since at the time of measurement the two
systems no longer interact, no real change can
take place in the second system in consequence
of anything that may be done to the first system.
This is, of course, merely a statement of what is
meant by the absence of an interaction between
the two systems \cite[p. 779]{eprP}.
 \end{quote}
 This statement must be compared with the one concerned with the {\em locality condition}:
 \begin{quote}\small \label{loccondition}
    If the measurement on part II of the system is made at the instant $t_{II}$ while that on part I was performed at the previous instant $t_{I}$ and $l>c(t_{II}-t_{I})$ -- where $l$ is the distance between the two points in which the two measurements are performed --  then the result obtained on part II can not be influenced by a physical interaction coming from part I.
 \end{quote}
 The locality condition as stated above, is implied by special relativity in which no physical interaction can propagate at a speed higher than that of light in vacuum.
 \par
The aim of the EPR paper was that of proving that quantum mechanics is an incomplete theory on the basis of a quite questionable definition of completeness. Moreover, the concept of `element of physical reality' used in the paper appears as ambiguous and, above all, superfluous. What is worst, this concept  has entangled in a spurious way philosophical considerations with technical ones concerning the completeness of a theory. This original sin of the EPR paper has never been redeemed: indeed, the spurious intertwining of philosophy and physics has been accentuated by some subsequent developments to the detriment of the
necessary and vital separation between physical theories and philosophical reflections on
 them.
\par
However,
  in spite of the flaws of the EPR paper, the thought experiment contained in it --
  opportunely reformulated -- has become the starting point of new physics and new applications. Some of these developments, will be discussed in the remaining sections.
 \section{Bohm's version of the EPR's thought experiment\label{bohmsec}}
In the Fifties, David Bohm tried to clarify some aspects of the EPR argument from two sides.
   Firstly, he reformulated the EPR thought experiment in the case of a system of two
particles with spin
 $\hbar/2$ in the singlet state \cite{bohm_1}. The two particles, are initially permitted to
interact
 (for instance, they are somehow produced by the same source); then, they fly
apart in
 opposite directions \footnote{The example discussed by Bohm was that of an excited Hydrogen molecule that splits into two hydrogen atoms.}. When the two particles  do not interact further, a
measurement of,
 say, the spin component of particle 1 along the $x$ direction is made. If, for
example,
 the result is $\hbar/2$, then we know with certainty that a measurement of the
spin component
  along the $x$ direction of particle 2 will yield $-\hbar/2$. Therefore, this
spin component of
   particle 2 is an element of physical reality. However, since we could have
measured,
   instead of the $x$ component of particle 1, the $y$ component, we would have
found that {\em also}
   the $y$ component of the spin of particle 2 is an element of physical
reality. Since,
   quantum mechanics can not attribute definite values to  both spin components of particle 2,
because they
    are incompatible quantities, quantum mechanics is incomplete.
    \par
 Secondly, in order to reintroduce a deterministic behavior of an individual
system,
 Bohm tried to reformulate the  formalism of non relativistic
  quantum mechanics by
 introducing `hidden' variables \cite{bohm_2, bohm_3}.   While the first step transformed the EPR
`thought
 experiment' in an experiment in principle realizable, the introduction of
hidden variables
   gave new fuel to the debate.
   \par
        Therefore, the appearance of a paper
   by John Bell was welcomed as a turning point: nothing would have been the
same as
   before \cite{bell64}.  Bell translated into a mathematical, general,  form the idea of
hidden
   variables   applied to the Bohm's version of the EPR thought experiment.
   He showed that an appropriately defined correlation coefficient of a
    hidden variables theory must satisfy  an inequality (Bell's inequality),
susceptible
     of an experimental check.
    Soon after,  Clauser, Horn, Shimony and Holt (CHSH)  generalized Bell's
    theorem  and proposed an experimental test based on the polarization
    correlations   of photon pairs  emitted by Calcium atoms  in a
    cascade process  \footnote{This proposed experiment was
    an extension of a previous one, performed by Kocher and Commins \cite{kocher}.} \cite{clauser}.   Aspect, Grangier and Roger realized the experiment proposed by
    CHSH using linear polarizers \cite{aspect81} and, soon after, two channels polarizers \cite{aspect82}. Both
    experiments strongly violated CHSH inequalities, showing that, in some
    circumstances, any hidden variables theory is disproved by experiment
    which, instead, corroborates the predictions of quantum mechanics. From
    then on, increasingly sophisticated experiments have been carried out in
    order to close several  loopholes: among them,  the communication
    loophole \cite{wheis}, due to the necessity of correlating polarization measurements
     space - like separated;   the detector loophole, due to the
    limited  efficiency of the photon detectors \cite{giustina}; and the coincidence loophole connected to the necessity of determining which local events form a pair   \cite{strong}. For a review of the loopholes issue, see, for instance, \cite{larsson}.
     \par
     In spite of the considerable amount of unequivocal experimental data, the debate about some EPR issues is not over.  In particular,
    I believe  that  two  questions require further critical inquiry:
    the role of the locality condition in hidden variables theories and  the interpretation of EPR - type experiments.
    In hidden variables theories, built up according to Bell - type theorems, the locality condition implies the statistical independence between the results of two measurements space - like separated:
 this implication leads hidden variables theories to be fatally disproved by experiment  (section \ref{belltype}).
On the other end,  in  standard interpretations of EPR - type experiments, a key role is played by the hypothesis that the twin photons of an entangled pair do not have  a definite polarization before measurement (NDV hypothesis) and by the  correlated hypothesis of non - locality (section \ref{orsaysec}).  These two hypotheses are used only for describing experiments and never for making predictions. In section \ref{howto}, it will be shown that the two hypotheses can be experimentally tested.
  \section{The role of locality in Bell - type theorems\label{belltype}}
        John Bell put it very clearly in the Introduction of his paper \footnote{It must be stressed that in Bell's paper  no philosophical element perturbs the logical chain of the argumentation.}:
\begin{quote}
It is the requirement of locality, or more precisely that the result of a measurement on one system be unaffected by operations on a distant system with which it has interacted in the past, that creates the essential difficulty \cite[p. 195]{bell64}.
\end{quote}
 In order to
clarify this point, it is necessary to revisit the deduction chain of a Bell - type theorem.   Consider a system constituted by two parts which having interacted in the past -- because, for example, somehow produced by the same source -- are now  separated by the distance $l$. Let us further suppose that  $l>c\Delta t$ where $\Delta t$  is the time interval that separates the two measurements carried out on the two parts of the system.
  It is not necessary
  to specify which measurements we are talking about. That stated, a Bell - type
  theorem can be formulated in the following way.
  \begin{center}
    Bell - type theorem
  \end{center}
       A hidden variables theory formulated on the basis of the following
assumptions:
             \begin{quote}
                \begin{enumerate}
                  \item [P1]  The system constituted by the two parts has
                      properties not   described by the formalism of
                      quantum mechanics. \label{Po1}
                  \item [P2]  These properties, symbolized by the parameter
                      $\lambda$, may vary from one measurement cycle to
                      another. Hence the opportunity of introducing a
                      normalized probability distribution  $\rho(\lambda)$
                      of the parameter $\lambda$. \label{Po2}
                      \item [P3] \label{Po3} The  locality condition (p. \pageref{loccondition})
                          {\em implies} that the results of the
                          measurements space - like separated performed on the two
                          parts of the system are statistically
                          independent.
\end{enumerate}
                                  \vskip1.5mm
                                  yields predictions that, in some circumstances,  are different
                from those of quantum mechanics.
                                  \end{quote}
                                   The assumptions P1 - P3 lead to the equation \footnote{See, for instance, \cite[p. 421]{cripto}.}:
    \begin{equation}\label{localend}
      p(ab|xy)= \int \rho(\lambda)
p(a|x,\lambda)p(b|y,\lambda)d\lambda
     \end{equation}
Where: the $p$'s are probabilities; $x$ and $y$ represent the types of measurement (not specified) made on part I and II of the pair;  $a$ and $b$ are the results of these measurements.                                           The statistical independence between the results $a$ and $b$ is represented in equation (\ref{localend})  by the product of the probabilities $p(a|x,\lambda)p(b|y,\lambda)$.  From equation (\ref{localend}) Bell - type inequalities  can be straightforwardly derived; am\-ong them, Bell and CHSH inequalities.  These inequalities must be satisfied by hidden variables theories built up according to Bell - type theorems. As  discussed in detail in section \ref{orsaysec}, these inequalities are violated by EPR  experiments which, on the other hand, corroborate the predictions of quantum mechanics.
\par
The vital role played by equation (\ref{localend}) in Bell - type theorems has been stressed by many. For instance, \cite{cripto} write:
\begin{quote}\small
    It is relatively frequent to see a paper claiming to ``disprove''
Bell's theorem or that a mistake in the derivation of Bell inequalities
has been found. However, {\em once one accepts the definition} (\ref{localend}), it is a
quite trivial mathematical theorem that this definition is incompatible
with certain quantum predictions \cite[p. 421]{cripto} (my italics).
\end{quote}
\par\noindent
  Equation (\ref{localend}) relies on assumptions P1, P2 and P3. EPR experiments show that the results of measurements performed on the separated parts of the systems are statistical dependent. Then, at least one of the assumptions is false. The most suspected  is P3, because it establishes explicitly the statistical independence of the two measurements on behalf of the locality condition.  Let us consider the
 following logical chain (the symbol $\Downarrow$ represents  logical
implication): \vskip3mm
 \begin{minipage}{11cm}\label{indistat}
\begin{center}
    A\,=\,Special relativity\\
    $\Downarrow$\\
   B\,=\, Locality condition\\
    $\Downarrow$\\
   C\,= \,Statistical independence of the measurements space like separated [$l>c(t_{II}-t_{I})$] performed on the two parts of the pair\\
\end{center}
\end{minipage}
\vskip3mm\noindent
  EPR experiments show that C is false.  If we believe  that special relativity is true, we must conclude that the implication B $\Rightarrow$ C is false, i.e. that the locality condition  does not implies the statistical independence of the two space like separated measurements.
As a matter of fact, the assumption (P3) that locality implies statistical independence is based on the {\em additional, implicit}  assumption that a statistical dependence between two events must be due to a physical interaction between them.  This additional hypothesis rules out  the possibility that the correlation between these two events might be due to an intrinsic property of the physical system under study. If P3 is false, the relevance of Bell - type theorems will be highly reduced \footnote{Of course, also P1 may be false: this amounts to say that it is not possible to introduce hidden variables in the formalism of quantum mechanics.}.  
 \section{The Orsay experiment\label{orsaysec}}
\begin{figure}[htb]
  \centering{
  \includegraphics[width=7 cm]{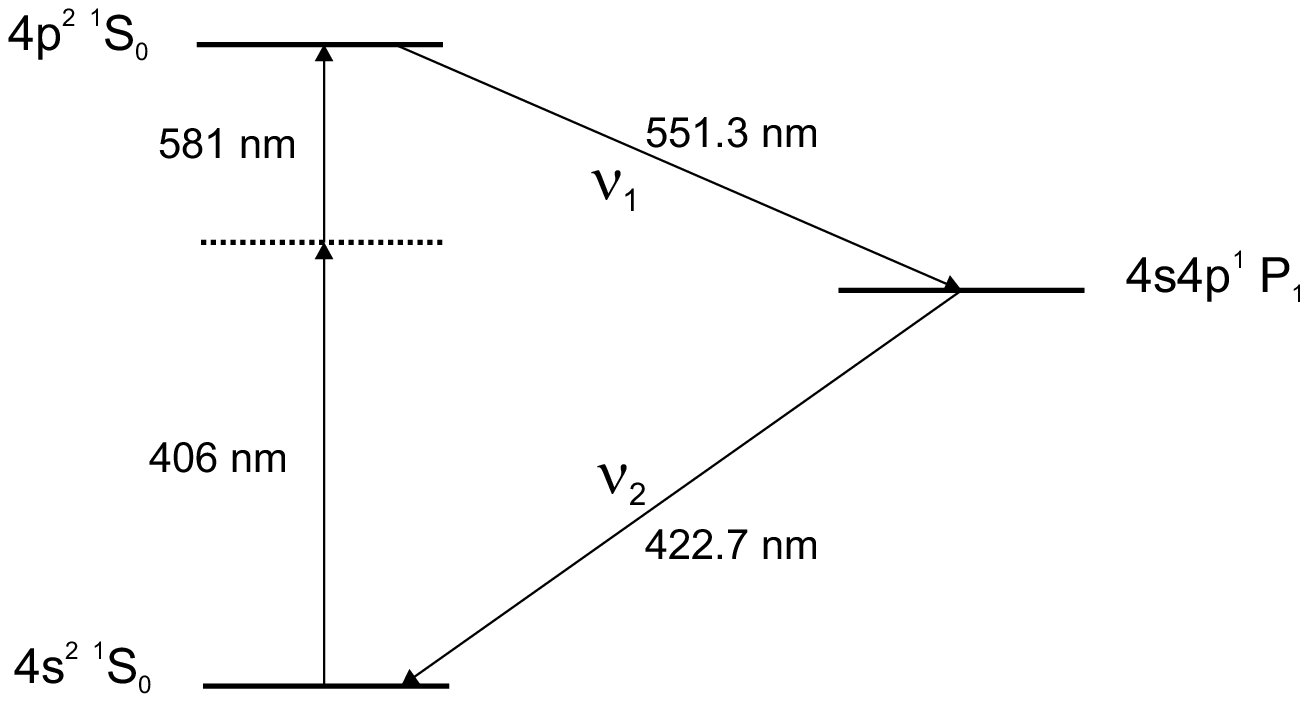}\\
  }
  \caption{Energy levels of Calcium atoms used in the Orsay experiment.}\label{livelli}
\end{figure}

In 1982, Aspect, Grangier and Roger carried out at Orsay a renowned EPR experiment \cite{aspect82}.
  The source is a beam of Calcium atoms. Two photon transitions produced by two laser beams, perpendicular to the atoms' beam, excite the electrons from the fundamental state  $4s^2$ ($J=0$) to the excited state $4p^2$ (J=0). From this state, the electrons fall back into the fundamental state by passing through the intermediate state $4s4p^1$ ($J=1$) and by emitting two photons corresponding to $\lambda_1=551.3\, nm$ and $\lambda_2=422.7\, nm$, respectively (fig. \ref{livelli}).
\par\noindent
 Let us consider the twin photons of a pair which propagate in opposite
directions $\pm z$. Two filters allow the propagation of  photons $\nu_1$ along $+z$ and of photons $\nu_2$ along $-z$. A two channels analyzer A, followed by two photomultipliers, is  situated along $+z$; an identical analyzer B  is situated along $-z$.
\begin{figure}[h]
  \centering{
  \includegraphics[width=10cm]{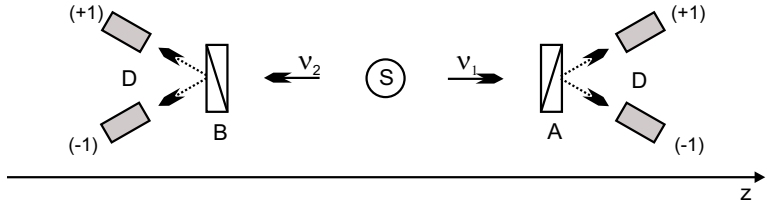}\\
  }
  \caption{Experimental setup of the Orsay experiment. $S$ is the source; A and B are two channels polarizers; D are photon detectors. Two filters, not shown in the figure, allow the passage of only photons $\nu_1$ along $+z$ and of photons $\nu_2$ along $-z$. See the text.}\label{eprfig}
\end{figure}
 \par\noindent
The experiment consists in measuring the polarization correlations of the twin photons as a function of the angle $\theta$ between the directions $\bf a$ and $\bf b$ that identify the orientations of the two polarizers.  The results are confronted with the predictions of quantum mechanics and those of  Bell - type hidden variables theories.
\par
 For the conservation of angular momentum in the emission process, the photons pairs are described by the state vector:
 \begin{equation}\label{circolare}
|\psi(\nu_1,\nu_2)\hspace{-2pt}>\:=\frac{1}{\sqrt{2}}\,(|R_1,R_2\hspace{-4pt}> +\,|L_1,L_2\hspace{-4pt}>)
\end{equation}
The state vector (\ref{circolare}) is an example of what are called entangled states. Since a circularly polarized photon can always be expressed in terms of two orthogonal linear polarizations, the photon pairs can be described by the state vector:
 \begin{equation}\label{corre}
|\psi(\nu_1,\nu_2)\hspace{-2pt}>\:= \frac{1}{\sqrt{2}}\,(|x_1,x_2\hspace{-4pt}>+\,|y_1,y_2\hspace{-4pt}>)
\end{equation}
 Let us suppose that the measurement on photon $\nu_1$ is made before that on photon $\nu_2$ and that the space - time interval separating the two measurements is space - like \footnote{Here, I am following the description given by Aspect \cite{aspectnaive}. In the Orsay experiment, the measurements on the  twin photons were not space - like separated. The first experiment of this kind, has been carried out by Weihs et al. in 1998 \cite{wheis}.}. The probability that  the photon $\nu_1$ enters one of the two channels of the analyzer A is $1/2$. If $\bf a$
is the polarization direction of photon $\nu_1$ {\em after the measurement}, then the photon pair, after the measurement made by A, is described by the state vector (reduction of the  state vector due to the measurement):
 \begin{equation}\label{dopoa}
    |\psi'(\nu_1,\nu_2)\hspace{-4pt}>= |\bf a,a\hspace{-4pt}>
\end{equation}
After the measurement by A,
 we {\em know} that photon $\nu_1$ is polarized along $\bf a$ (because it has come out from a polarizer oriented along $\bf a$), and we  {\em learn}, from the state vector reduction, that the polarization of photon $\nu_2$ is the same as that measured for $\nu_1$.
Therefore, if the analyzer B is oriented as A (i.e. if the direction $\bf a$ and $\bf b$ are parallel), the photon $\nu_2$ will pass through,  with certainty, the same  channel
 passed through by $\nu_1$. If, instead, $\theta$ is the angle between $\bf a$ and $\bf b$, photon $\nu_2$ will pass through the same channel passed through by photon $\nu_1$ with probability $\cos^2\theta$ (Malus law). Hence,
 the probability that $\nu_1$ and $\nu_2$ pass through the same channel and, then, will have  parallel  polarizations after the measurements, is given by:
\begin{equation}\label{qmab}
    P_{\|\|}({\bf a},{\bf b})=\frac{1}{2}\cos^2\theta
\end{equation}
This equation can, of course, be derived directly from (\ref{corre})
 without considering the
details of the two steps description.
 Analogously,  we can derive   the probability that the two photons pass through  different channels:
\begin{equation}\label{qmabper}
    P_{\|\perp}({\bf a},{\bf b})=\frac{1}{2}\sin^2\theta
\end{equation}
These equations, together with similar ones   for  $P_{\perp\perp}$ and $P_{\perp\|}$, are used to calculate the quantum mechanical value of the quantity $S$ appearing in the CHSH inequality. The quantity $S$ is defined as:
\begin{equation} S= E(\bf a, b)- E(\bf a, b')+E(\bf a',b)+ E (\bf a',b')
 \end{equation}
 where the $E$'s are the correlation coefficients corresponding to  the pairs of orientations of the two analyzers A and B. We recall that, for example:
\begin{equation}
E({\bf a, b})= P_{\|\|}({\bf a, b})+P_{\perp\perp}({\bf a, b})-P_{\|\perp}({\bf a, b})- P_{\perp\|}({\bf a, b})=\cos 2({\bf a, b})
\end{equation}
  The Orsay experiment shows that the maximum violation of the CHSH inequality occurs in correspondence of a value of  $S_{QM}=\pm 2.697 \pm 0.015$. This value must be compared with the theoretical one $\pm 2\sqrt{2}=\pm 2.8284$ and the fact that, according to hidden variables theories $-2\le S \le 2$. Therefore, the Orsay experiment shows that hidden variables theories can not fully reproduce the predictions of quantum mechanics and that, on the other hand, these predictions are in satisfactory agreement with experiment.
\par
 In the above two steps description of the  Orsay experiment two  points are crucial:
 \begin{enumerate}
   \item[\fbox{I}] No  supposition is made about the value of the
       polarization of photon $\nu_1$ before its measurement. The reason is
       that, for using the state vector reduction due to the measurement,
       only the value of the polarization after the measurement is of
       interest.  \label{nosup}
    \item[\fbox{II}] The reduction of the state vector due to the
        measurement on  photon $\nu_1$, shows that the polarizations of the
        twin photons are strongly correlated: see equation (\ref{dopoa}).
         \label{shows}
        \end{enumerate}
  This description -- which we shall denote as the `formal description' -- uses only the formalism of quantum mechanics without any further assumption. It asserts only that the measurement implies -- at the theoretical level -- a reduction of the state vector of the photon pair and that this reduction uncovers the polarization correlation between the twin photons. Then, on the basis of the new state vector, it predicts which will be the result of the polarization measurement on photon $\nu_2$. This description follows strictly the prescriptions of quantum mechanics: i) write the state vector for the photon pairs on the basis of the production (preparation) process of the twin photons; ii) make predictions by using the formalism; iii) compare the predictions with the experimental results.
 \par
Instead, the standard description uses  two additional hypothesis \footnote{See, for instance, the paper by Aspect \cite{aspectnaive}.}:
\begin{enumerate}
  \item[\fbox{A}] The twin photons do not have a definite polarization
      before measurement (hypothesis NDV (Not Definite Value)). \label{ndv}
  \item[\fbox{B}] When the measurement on photon $\nu_1$ is made, photon
      $\nu_2$ assumes instantaneously the same polarization of photon
      $\nu_1$.\label{spooky}
 \end{enumerate}
  If we want to describe the outcomes of the experiments on the basis of hypothesis \fbox{A}, we must necessarily adopt also hypothesis \fbox{B}.
As we have seen in point \fbox{I}, the polarization of  photon $\nu_1$ before the measurement is not used in  the calculation of the quantum mechanical predictions. Hence, the NDV hypothesis is used only for describing, a posteriori, the predictions obtained by \fbox{I} and \fbox{II}. Anyway, in next section, it is shown how the \fbox{A} (NDV) and \fbox{B} hypotheses can be tested by experiment.
\par
Statement \fbox{B}\,  has the same logical structure of the statement: when I push the button,  the lamp  turns on. The statement on the lamp implies a causal connection between the pushing of the button and the turning on of the lamp. This causal connection is suggested by the detailed knowledge of the physical system to which the button and the lamp belong. Also the statement about photon $\nu_2$ should contain a similar causal connection. However, a causal connection is forbidden by the fact that the two measurements on the twin photons are space - like separated. Therefore, it is necessary to invoke a kind of instantaneous action at a distance between the two  measurements. This action at a distance is labeled as non - locality. However, non - locality is used only for describing the experiment and not for making predictions: these are obtained only from the formalism of quantum mechanics. Therefore, the concept of non - locality can be dropped without reducing the predictive power of quantum mechanics concerning entangled pairs.
\par
It is worth stressing that the description of the experiment based on the NDV and the non - locality hypotheses presupposes that the things go in the World, step by step, exactly as described by the theory. Therefore, this description is a paradigmatic example of what is called realism of theories, a type of realism that appears as untenable (section \ref{philback}).
\par
Finally, the standard description of the Orsay experiment faces what may be called the simultaneity paradox. Let us suppose that the two measurements are made at the same instant.  If $\nu_1$  enters channel ($+1$), then $\nu_2$ enters simultaneously the same channel. However, since the two measurements are simultaneous, $\nu_2$ might enter channel ($-1$): in this case, $\nu_1$ should enter the same channel, in contradiction with the starting hypothesis.
\section{How to experimentally test the NDV and the non - locality hypotheses\label{howto}}
The NDV hypothesis is strictly connected to that of non - locality: both are used in the standard description of the Orsay experiment or, more generally, of EPR - type experiments. The NDV and the non - locality hypothesis can be experimentally tested by a modification of the apparatus of the Orsay experiment.
      \par
During the emission of each photon of a pair, the total angular momentum (atom + emitted photon) must be conserved. Therefore, the twin photons emitted in a single cascade process should be both right ($R$) or left ($L$) circularly polarized. This implies that the photons of each pairs have definite polarization when emitted by the Calcium atoms, i.e. {\em before} any polarization measurement. This assertion is the negation of  the NDV (Not Definite Value) hypothesis (the twin photons do not have a definite polarization before measurement) that, instead, is used in the standard description  of the Orsay experiment. Luckily, there is an experimental procedure that allows to verify which is the polarization of a light beam {\em before} the measurement. The procedure requires the use of a quarter wavelength plate and a linear polarizer and it  can be applied to the twin photons emitted by  Calcium atoms \footnote{A similar experiment has been already proposed in a previous work of mine \cite[p. 273]{ggnc}.}. In an Orsay - type apparatus, the double channel analyzers must be replaced, in both arms, by a quarter wavelength plate and a linear polarizer. We shall suppose that the two measurements on photon $\nu_1$ and photon $\nu_2$ are made at two  instants such that the space - time interval separating the two measurements is space - like. The quarter wavelength plate transforms the incoming photon supposed to be $R$ ($L$) polarized  in a photon linearly polarized along a direction which makes an angle of $\pi/4$ ($-\pi/4$) with the optical axis of the plate \footnote{Here, we are adopting the convention according to which a photon is $R$ ($L$) polarized if its intrinsic angular momentum vector is directed along (opposite to) its propagation direction. }. Therefore, if after the plate, we insert a linear polarizer whose optical axis is parallel to the direction $\pi/4$ ($-\pi/4$), the photomultiplier situated  after the polarizer will click if the incoming photon is $R$ ($L$) as emitted by the source. If in the two arms of the apparatus the linear polarizers are both oriented along a direction which makes an angle of $\pi/4$ ($-\pi/4$)  with the optical axis of the associated plate, both photomultipliers placed behind the linear polarizers will click only if the twin photons are produced as $R$ ($L$) circularly polarized by the source, i.e. before the polarization measurements.
 \par
 Now, let us try to describe this proposed experiment on the basis of hypothesis \fbox{A} (NDV hypothesis) and of hypothesis \fbox{B} (non - locality hypothesis).
The NDV hypothesis affirms that photon $\nu_1$ does not have a definite polarization before measurement. However, if  photon $\nu_1$ is detected, this photon, after the measurement of its polarization (i.e. after the linear polarizer) is linearly polarized along the direction of the optical axis of the polarizer. Let us denote this direction with the symbol $\|$. According to the non - locality hypothesis, photon $\nu_2$ assumes instantaneously the same linear polarization  $\|$. Then, photon $\nu_2$, going through the plate, will be transformed into an $R$ circularly polarized photon and will have a probability $1/2$ of going through the subsequent linear polarizer. Therefore: while the hypothesis that the twin photons are both produced as $R$ or $L$ polarized yields a probability of being detected for photon $\nu_2$ equal to one, the NDV and non - locality hypotheses yield for the same probability the value of $1/2$. Then, the proposed experiment allows to experimentally test the two theoretical predictions and shows that
  the question of the polarization of the twin photons as produced by the source in a physical question, answerable with an experiment.
 \section{Discussion\label{then}}
 Initially, EPR experiments have been devised in order to disprove hidden variables theories (built up as indicated by Bell - type theorems). EPR experiments have disproved these theories and   ascertained  that the polarizations of twin photons  described by entangled states obey quantum mechanics. But this is only a  piece of the story.  In fact, these experiments have taught us how to cleverly  manipulate  entangled pairs of photons and basically contributed to new fields of applied research like that of quantum information and quantum cryptography \cite{cripto}. These applications depend only on the correlation properties of entangled states. The validity of Bell - type theorems  or the way in which we describe EPR experiments are irrelevant for these applications. These questions are instead of interest from an epistemological viewpoint. The following considerations are dedicated to these issues.
\par
Typical  consequences usually drawn from EPR experiments are:
\begin{enumerate}
  \item Hidden variables theories formulated according to Bell - type
      theorems are disproved by experiments \label{hvex}
  \item Quantum mechanics is non - local  \label{nonlocal}
    \item Realism and locality are incompatible \label{realloc}
\end{enumerate}
 The  epistemological status of these
  statements is  different.
   Statements \ref{hvex} and \ref{nonlocal} are physical statements, while statement \ref{realloc} mixes up physical and  philosophical issues.
\par
  As we have seen,  statement \ref{hvex} is true. However,  hidden variables theories considered by Bell - type theorems are built up with an intrinsic lethal  flaw, constituted by equation (\ref{localend}): this equation is a direct consequence of the  implication `locality condition $\rightarrow$ statistical independence of the results of two space like separated measurements'. This implication is valid only if it is additionally assumed that statistical dependence between two events requires a physical interaction between them. This additional assumption forbid hidden variables theories to describe statistical dependencies due to intrinsic properties of the physical system. Therefore,  hidden variables theories are built up with rules that fatally lead them to be disproved by experiments: they are `straw man' theories.
\par
Statement \ref{nonlocal} follows from statement \ref{hvex} if the implication `locality $\Rightarrow$ statistical independence of the results of two measurements  space like separated' is true. However, the logical chain considered in section \ref{belltype} shows that  this implication is false (if we believe that special relativity is true) and, therefore statement \ref{nonlocal} appears as questionable. However, the experiment proposed in the previous section allows a direct experimental test of non - locality.
\par
 Statement \ref{realloc}
  entangles physical elements (locality)    with philosophical ones  (realism), following the  temper of the EPR paper. It
  has been phrased in many ways.
  Here are some representative examples:
\begin{quote}
    The violation of a Bell inequality is an experimental observation that forces the abandonment of a local realistic viewpoint -- namely, one in which physical properties are (probabilistically) defined before and independently of measurement, and in
which no physical influence can propagate faster than the speed of light \cite[p. 227]{giustina}. \vskip3mm
\par
 Bell's theorem does not prove the validity of quantum
mechanics, but it does allow us to test the hypothesis that nature is governed by local realism. The principle of realism says that any system has preexisting values for all possible measurements of the system. In local realistic theories, these preexisting values depend only on events in the past light cone of the system \cite[p. 250402 - 2]{strong}. \vskip3mm\par
 Most working scientists hold fast to the concept of `realism' -- a viewpoint according to which an external reality exists
independent of observation. But quantum physics has shattered some of our cornerstone beliefs. According to Bell's theorem, any theory that is based on the joint assumption of realism and locality (meaning that local events cannot be affected by actions in space - like separated regions) is at variance with certain quantum predictions. Experiments with entangled pairs of particles have amply confirmed these quantum predictions, thus rendering local realistic theories untenable. Maintaining realism as a fundamental concept would therefore necessitate the introduction of `spooky' actions that defy locality \cite[p. 871]{nonlocalcit}.
\end{quote}
 The meaning of the word `realism'  is not the same in these quotations. In the first two,
by `realism' it is meant that a physical quantity has a definite value before measurement. It is true that this assumption is suggested by a philosophical realistic stand. However, as shown in section \ref{howto}, the assertion that the photons of an entangled  pair have or not have a definite value of polarization before measurement is a physical question, answerable by an experiment. The experiment tests a physical assertion and not the philosophy that might have inspired it.
 \par
 In the third quotation, by realism it is meant  `a viewpoint according to which an external reality exists
independent of observation.' This philosophical viewpoint has been  at the basis of the scientific endeavor and of what we may call the rationally oriented common sense. This philosophical stand (like any other) can not be falsified by any experiment. As already pointed out in section \ref{philback}, in order to experimentally disprove a philosophy, one should  build up a theory whose entire set of assumptions are logically derived from the philosophy and shows that this theory is falsified by experiment.
 \par
However, the philosophical stand of a scientist in not a neutral one with respect to his attitude in elaborating theories, devising or explaining experiments. Different philosophies  produce different heuristics.    For instance, a realistic stand will suggest to search for the causes of a phenomenon, or, more specifically, to investigate further the polarization properties of a photon in an EPR experiment in order to better understand its behavior. Instead, other philosophies will hold that the probabilistic nature of quantum mechanics reflects the indeterministic behavior of the microscopic world. Consequently, it is useless to search for causes or, more specifically, to investigate further the polarization properties of the photons of an entangled pair. The difference between these two heuristics showed up in the discussion of the role of locality in Bell - type theorems (section \ref{belltype}) and of the Orsay experiment (section \ref{orsaysec}) or in the proposal of experiments for testing the NDV and the non - locality hypothesis (section \ref{howto}).
\section{Conclusions}
A critical reconsideration the EPR (Einstein - Podolsky - Rosen) paper shows  that
   the  EPR argument can be developed without using the
      concept of `element of physical reality', thus eliminating any
      philosophical element in the logical chains of the paper.
      Deprived of its philosophical ornament, the EPR argument plainly
      reduces to require what quantum mechanics can not do: to assign
      definite values to two incompatible physical quantities. However, the thought experiment devised by EPR, opportunely revised, has become a laboratory experiment and has given birth to  new fields of fundamental ad applied research.
      \par
      Hidden variables theories built up according to Bell - type theorems are formulated on the basis of  the assumption that the locality condition implies the statistical independence between two measurements space - like separated.  This assumption is valid only with the additional one that statistical dependence between two measurements requires a physical interaction between them. This additional assumption rules out the possibility that  statistical dependence may be due to an intrinsic property of the physical system under study. Therefore, hidden variables theories are built up with a  restriction which leads them to be disproved by experiment. They appear to be `straw man' theories whose main role seems that of allowing to call into question philosophical realism. However, a philosophy can not be disproved by an experiment unless it is shown that this experiment disproves a  theory whose postulates are  logical consequences of the philosophy.
  \par
   Quantum mechanical non - locality, invoked for describing EPR -
      type experiments, is strictly connected to the hypothesis (NDV hypothesis) according to which the twin photons of entangled pairs do not have a definite polarization before measurements. Both hypotheses are used only for describing EPR experiments and not for making predictions. Therefore, they can be dropped without reducing the predictive power of quantum mechanics concerning entangled photons pairs.  Furthermore,  both hypotheses
      can be experimentally tested by a  modification of a standard experimental
      apparatus designed for studying entangled photons pairs.
      \vskip3mm\par
            {\bf Acknowledgements.} I would like to thank Biagio Buonaura who has sympathetically and critically followed, step by step, the writing of this paper.


\begin{thebibliography}{}
\bibitem{eprP} Einstein A, Podolsky B and Rosen N {1935} Can
Quantum - Mechanical
      Description of Physical Reality Be Considered Complete?
     {\em Phys. Rev.}
     \href{https://journals.aps.org/pr/abstract/10.1103/PhysRev.47.777} {{\bf 47} {777-780}}.
     \bibitem{cripto}   Brunner N, Cavalcanti D, Pironio S, Scarani V and
Wehner S 2014 Bell nonlocality {\em Rev. Mod. Phys.}
\href{https://journals.aps.org/rmp/abstract/10.1103/RevModPhys.86.419} {{\bf
86} 419-478} \href{https://arxiv.org/abs/1303.2849}{arxiv}.
\bibitem{erq} Giuliani G 2019 \href{http://www.paviauniversitypress.it/catalogo/elettromagnetismo-relativita-quanti/2588}{\em Elettromagnetismo, relativit\`a, quanti - Fisica, storia, epistemologia} (Pavia, Pavia University Press).
  \bibitem{etere}  Maxwell J C, 1879 `Ether' in:  {\em Enciclop{\ae}dia Britannica} IX
                    edn    vol 8.  (Encyclop{\ae}dia
Britannica Inc., Edinburgh) online \href{http://ether-wind.narod.ru/Maxwell_Britannica_Ether_1878/Maxwell_Britannica_Ether_1878.pdf}{here}.
\bibitem{what} Giuliani G 1998 What are physicists talking about?
The case of electrons and holes {\em  N. Cim.} D \href{https://link.springer.com/article/10.1007%2FBF03185528}{{\bf 20} 1183-1186}.
     \bibitem{bohm_1}   Bohm D, 1951 {\em Quantum Theory}. (Prentice Hall, New York).
         \bibitem{bohm_2}Bohm D
1952 A Suggested Interpretation of the Quantum Theory in Terms of ``Hidden''
variables I  {\em Phys. Rev}
\href{https://journals.aps.org/pr/abstract/10.1103/PhysRev.85.166}{{\bf 85}
166-179}.
          \bibitem{bohm_3}   Bohm D 1952 A Suggested Interpretation of the Quantum Theory
in Terms of ``Hidden'' variables II  {\em Phys. Rev}
\href{https://journals.aps.org/pr/abstract/10.1103/PhysRev.85.180} {{\bf 85}
180-193}.
 \bibitem{bell64} {Bell J} 1964 On the Einstein Podolsky Rosen paradox \href{https://cds.cern.ch/record/111654/files/vol1p195-200_001.pdf}{{\em Phys.}
    {\bf 1} {195-200}}.
  \bibitem{clauser}  Clauser J, Horne M, Shimony A and Holt R 1969 Proposed
    experiment to test local hidden - variables theories {\em Phys. Rev. Lett.}
    \href{https://journals.aps.org/prl/abstract/10.1103/PhysRevLett.23.8800000000}
    {{\bf 23} 880-884}.
       \bibitem{kocher}   Kocher C and Commins E 1967 Polarization correlations of photons
emitted in an atomic cascade  {\em Phys. Rev. Lett.}
\href{https://journals.aps.org/prl/abstract/10.1103/PhysRevLett.18.575} {{\bf
18} 575-577}. \bibitem{aspect81}  Aspect A,  Grangier P and   Roger G 1981  Experimental Tests
of Realistic Local Theories via Bell's Theorem
      {\em Phys. Rev.
     Lett.} \href{https://journals.aps.org/prl/pdf/10.1103/PhysRevLett.47.460}{{\bf 47} 460-463}.
             \bibitem{aspect82}  Aspect A,  Grangier P and   Roger G 1982  Experimental
Realization of Einstein-Podolsky-Rosen-Bohm {\em Gedankenexperiment}: A New
Violation of Bell's Inequalities {\em Phys. Rev.
     Lett.}
     \href{https://journals.aps.org/prl/abstract/10.1103/PhysRevLett.49.91}
     {{\bf 49} {91-94}}.
\bibitem{wheis}  Weihs G,   Jennewein T,   Simon C,   Weinfurter H and  Zeilinger A 1988 Violation of Bell's Inequality under Strict Einstein Locality Conditions  {\em Phys. Rev. Lett.}, \href{https://journals.aps.org/prl/pdf/10.1103/PhysRevLett.81.5039}{{\bf 81} 5039-5043}.
\bibitem{giustina}   Giustina M et al. 2013 Bell violation using entangled
photons without the fair - sampling assumption {\em Nature}
\href{https://www.nature.com/articles/nature12012} {{\bf 497} 227-230}
\href{https://arxiv.org/abs/1212.0533}{arxiv}.
              \bibitem{strong} Shalm L K et al.
2015 Strong Loophole - Free Test of Local Realism {\em Phys.Rev. Lett.}
\href{https://journals.aps.org/prl/abstract/10.1103/PhysRevLett.115.250402}
{{\bf 115} 250402} \href{https://arxiv.org/abs/1511.03189}{arxiv}.
\bibitem{larsson}
   Larsson J {\AA}, 2014 Loopholes in Bell inequality tests of local realism {\em Jour. Phys.} A, \href{https://iopscience.iop.org/article/10.1088/1751-8113/47/42/424003/pdf}{{\bf 47}  424003} \href{https://arxiv.org/pdf/1407.0363.pdf} {arxiv}.
     \bibitem{aspectnaive} Aspect A 2002 Bell's theorem: the naive view  of an
experimentalist in: Bertlmann R and Zeilinger A (eds.) {\em Quantum
(Un)speakables} (Springer, Berlin)
\href{https://arxiv.org/ftp/quant-ph/papers/0402/0402001.pdf}{arxiv}.
         \bibitem{ggnc}  Giuliani G 2007 On realism and quantum mechanics {\em N.
Cim.} B \href{https://www.sif.it/riviste/sif/ncb/econtents/2007/122/03/article/7}{{\bf 122} 267-276}.
         \bibitem{rosen}  Rosenfeld W,  Burchardt D,   Garthoff R,   Redeker K,
 Ortegel N,   Rau M and   Weinfurter H, 2017 Event - Ready Bell Test Using Entangled Atoms Simultaneously Closing Detection and Locality Loopholes {\em Phys. Rev. Lett.}, \href{https://journals.aps.org/prl/abstract/10.1103/PhysRevLett.119.010402}{{\bf 119} 010402}.
         \bibitem{nonlocalcit}  Gr\"oblacher S,  Paterek T,  Kaltenbaek
    R,  Brukner \v{C}, Z\"{y}ukowski M,
 Aspelmeyer M and  Zeilinger A 2007 An experimental test of non - local realism
{\em Nature} \href{https://www.nature.com/articles/nature05677} {{\bf 446} 871-875}
\href{https://arxiv.org/abs/0704.2529}{arxiv}.

\end{thebibliography}
\end{document}